\documentstyle[aps,prl,twocolumn,epsf]{revtex}
\begin{document}
\draft
\title{ Random matrix model for quantum dots with interactions and 
the conductance peak spacing distribution}

\author{ Y. Alhassid, Ph. Jacquod and A. Wobst}

\address{ Center for Theoretical Physics, Sloane Physics Laboratory, Yale
University, New Haven, Connecticut 06520, USA}

\date {\today}

\maketitle
\begin{abstract}
 We introduce a random interaction matrix model (RIMM) for finite-size 
strongly interacting
fermionic systems whose single-particle dynamics is chaotic.
The model is applied to Coulomb blockade quantum dots with 
irregular shape to describe the crossover of the
peak spacing distribution from a Wigner-Dyson to a Gaussian-like
distribution. The crossover is universal within the random matrix
 model and is shown to depend on a single
 parameter: a scaled fluctuation width of the interaction matrix elements. The crossover observed in
 the RIMM is compared with the results of  an Anderson model with Coulomb 
interactions.

\end{abstract}


\narrowtext

 The transport properties of semiconductor quantum dots 
can be measured by connecting the dots to leads via point contacts
\cite{Kastner92}. When these point contacts are pinched off, 
effective barriers are  formed between the dot and the leads, and the charge
 on the dot is quantized.
Adding an electron to the dot requires a charging energy $E_C$ to
 overcome the Coulomb repulsion
with electrons already in the dot.
This repulsion can be compensated by
modifying the gate voltage $V_g$ on the dot. For temperatures below $E_C$, a
series  of Coulomb blockade oscillations is
observed in the linear conductance as a function of $V_g$. For
temperatures much smaller than the mean level spacing $\Delta$, the
conductance  is dominated by resonant tunneling and the Coulomb blockade
oscillations  become a series of sharp peaks.

 In dots with irregular shapes, the classical single-electron
 dynamics is mostly chaotic.  Quantum mechanically, chaotic systems 
are expected to exhibit universal fluctuations that are
described by random matrix theory (RMT). The distributions \cite{Jalabert92}
and parametric correlations \cite{Alhassid96} of the Coulomb blockade peak
heights in quantum dots have been derived using RMT, and these 
predictions have been confirmed experimentally \cite{Chang96,Folk96}.

 Another quantity of recent experimental and theoretical interest is the peak
spacing statistics. The peak spacing $\Delta_2$ can
be expressed as a second order difference of the ground state energy
${\cal E}_{\rm g.s.}^{(n)}$ of the $n$-electron
dot as a function of the number of electrons:
\begin{eqnarray}\label{Delta2}
\Delta_2 = {\cal E}_{\rm g.s.}^{(n+1)} + {\cal E}_{\rm g.s.}^{(n-1)} -2{\cal
E}_{\rm g.s.}^{(n)}
\;.
\end{eqnarray}
Using the constant interaction model (which ignores
interactions except for a classical Coulomb energy of $n^2 E_C/2$), and
assuming a single-particle spectrum that is independent of $n$,
$\Delta_2 = E_{n+1} -
E_n + E_C$, where $E_n$ is the $n$-th single-particle energy. Within 
this model, RMT suggests a Wigner-Dyson distribution of the peak spacings with a
 width of $\sim \Delta/2$. However recent experiments find a distribution 
that is Gaussian-like and has a larger width
\cite{Sivan96,Simmel97,Patel98,Simmel99}. This observation underlines
 the limitations of the  constant interaction model and the importance of
electron-electron interactions beyond an average Coulomb energy.
Some observed features of the peak spacing distribution have been reproduced
using  exact numerical diagonalization of small disordered dots ($n \alt 10$)
with  Coulomb interactions \cite{Sivan96,Berkovits98}. The width of the
distribution  is found to increase monotonically with the gas parameter $r_s$.
  Analytic RPA estimates in a disordered dot for small values of $r_s$ give
peak spacing fluctuations that are larger than those of RMT but still of
the  order of $\Delta$ \cite{Blanter97}. Recent Hartree-Fock
calculations \cite{Levit99,Walker99} of larger disordered and chaotic
dots with interactions (up to $n \sim 50$ electrons) also reproduce
 Gaussian-like peak spacing distributions.

 The above studies were carried out for particular models of quantum dots
 using Coulomb as well as nearest-neighbor interactions, and it is not clear 
how generic the conclusions are. It is also not obvious which bare
 electron-electron interaction should be taken to represent the experiments 
because of screening generated by external charges.   It is therefore 
important to find out whether the observed statistics of the peak spacings
 is  generic, and in particular whether it can be reproduced by a modified 
random matrix model. Standard RMT does not make explicit reference 
to interactions or to
number of particles. To study generic interaction effects on
the statistics, we need a random matrix model in which the two-body
interactions are distinguished from the one-body part of the Hamiltonian.
 Recently a two-body random
interaction model (TBRIM) introduced in nuclear physics
 \cite{French70} was used together with a
diagonal random one-body Hamiltonian to study thermalization in finite-size
systems \cite{Flambaum96} and the crossover from Poisson to
Wigner-Dyson statistics in many-body systems \cite{Jacquod97}. The
model explains statistical features observed in atomic \cite{Flambaum94} and
nuclear  \cite{Zelevinsky96} shell model calculations. However, the
 Poisson statistics that was used as a non-interacting limit of the
 model \cite{Jacquod97,Flambaum94} is
not  suitable for the study of dots whose single-electron dynamics is chaotic.

In this paper we introduce a random interaction matrix model (RIMM) for
strongly interacting Fermi systems whose single-particle dynamics is chaotic.
With this model we can study generic and universal effects associated with
the  interplay of one-body chaos and two-body interactions. In particular, we
apply
the model to study the peak spacing statistics and find a crossover from a
Wigner-Dyson  distribution to a Gaussian-like distribution as a function of
a parameter that measures the fluctuations of the interaction 
matrix elements. The crossover depends on both the number of
particles and the number of single-particle orbits but becomes universal upon
an  appropriate scaling of the interaction strength. The
crossover is demonstrated in a model of a small disordered dot with Coulomb
interactions, and we show that the results can be scaled approximately onto
those of the RIMM.

 A general Hamiltonian for spinless interacting fermions has the form
\begin{eqnarray}\label{Hamiltonian}
H = \sum\limits_{ij} h_{ij} a^\dagger_i a_j +{1\over 4} \sum_{ijkl}
\bar u_{ijkl}a^\dagger_i a^\dagger_j a_l a_k
\;,
\end{eqnarray}
where $h_{ij}$ are the matrix elements of the one-body Hamiltonian and $\bar
u_{ij;kl}  = \langle i j | u | k l \rangle - \langle i j | u| l k \rangle$ are
the  antisymmetrized matrix elements of the two-body interaction. The states
$|i\rangle  = a_i^\dagger |0\rangle$ describe a fixed basis of $m$
single-particle states. We define an ensemble of random matrices $H$ of the
form  (\ref{Hamiltonian}), where the one-body $m\times m$ matrix
$h_{ij}$ belongs to the Gaussian ensemble of symmetry class $\beta$, and the
matrix  elements of the two-body interaction are real independent Gaussian
variables  with zero average and variance $U^2$ ($\frac{1}{2}U^2$)
for the diagonal (off-diagonal)  elements
\begin{eqnarray}\label{random-ensemble}
P(h) \propto e^{-{\beta\over 2a^2} {\rm Tr}\; h^2} \;;\;\;\; P(\bar u)
\propto   e^{- {\rm Tr}\;\bar u^2/2U^2}
\;,
\end{eqnarray}
Eqs. (\ref{Hamiltonian}) and (\ref{random-ensemble}) define the RIMM. 
 The parameter $a$ determines the single-particle level spacing $\Delta$.
In the non-interacting  limit $U=0$, the random ensemble describes the
universal statistical properties of a system whose single-particle dynamics
is  chaotic. For conserved time-reversal symmetry $h$ is
a GOE matrix, while for broken time-reversal symmetry,
e.g. in presence of an external magnetic field,
$h$ becomes a GUE matrix.  The random
ensemble  for the two-body part of the Hamiltonian (\ref{Hamiltonian}) is
invariant  under orthogonal transformations of the single-particle basis.
An average interaction that is invariant under such transformations 
can be included in the model, but it leads to a constant charging energy
 shift in $\Delta_2$ and does not affect the peak spacing statistics.

 The randomness of the two-body interaction matrix elements is  motivated by
the strong fluctuations of the Coulomb interaction matrix elements in the
basis  of eigenstates of the chaotic single-particle Hamiltonian.
 The RIMM differs from the TBRIM \cite{Flambaum96,Jacquod97}
in its one-body part, which is less relevant at the
high excitation energies considered in the earlier studies but is of crucial
importance  near the ground state. The TBRIM has a Poissonian statistics in
the  non-interacting limit, in contrast to the Wigner-Dyson statistics
characterizing the non-interacting limit of (\ref{Hamiltonian}). Our
 random interaction
matrix  model is suitable for describing the generic statistical fluctuations
in quantum dots with chaotic single-particle dynamics and in the presence of
electron-electron  interactions. The model depends
on three parameters: $U/\Delta$, the number of single-particle orbits $m$,
 and the number of particles $n$.

\begin{figure}

\epsfxsize=3.3in
\epsfysize=2.2in
\epsffile{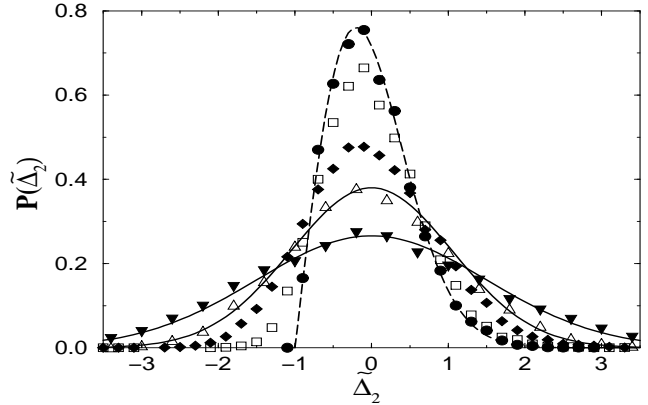}

\caption
{ Peak spacing distributions $P(\tilde \Delta_2)$ for the random matrix
model (\protect\ref{Hamiltonian}) for $m=12$, $n=4$ and for
$U/\Delta=0$ (solid circles), 0.35 (open squares), 0.7 (solid diamonds),
1.1 (open triangles) and 1.8 (solid triangles). The one-body part $h$ is a
GOE.  We see a crossover from a
Wigner-Dyson distribution for $U =0$ (dashed line) to
Gaussian-like distributions for $U/\Delta \protect\agt 1$. The solid lines are
Gaussian distributions with the widths of the $U/\Delta=1.1$ and $1.8$
distributions, respectively.
}
\label{fig:crossover}

\end{figure}

 Next we apply the RIMM(\ref{Hamiltonian}) to study the peak
spacing  statistics in Coulomb blockade quantum dots. Peak spacings are
 computed using (\ref{Delta2}); i.e. 
the ground-state energy is calculated for three
consecutive numbers of particles $n-1$, $n$ and $n+1$, and statistics
 are collected by generating
realizations of the ensemble $H$. 

 Typical distributions of $\tilde \Delta_2 \equiv (\Delta_2 - \langle
\Delta_2\rangle)/\Delta$ for the case of conserved
time-reversal symmetry ($h$ is a GOE
matrix) are shown in Fig. \ref{fig:crossover} for several values of
$U/\Delta$. For the non-interacting case we obtain the
Wigner-Dyson distribution (dashed line), but as $U/\Delta$ 
increases a crossover is observed to a Gaussian-like distribution. The
distributions for $U/\Delta=1.1$ and 1.8 are
compared with a Gaussian of the same width
(solid lines). The model (\ref{Hamiltonian}) does not 
include spin and therefore cannot reproduce the expected
 bimodal structure of the peak spacing distribution at weak
 interactions. However, numerical simulations 
in small disordered dots indicate that this bimodal structure
 disappears already for weak interactions \cite{Berkovits98}.

\begin{figure}

\epsfxsize=3.3in
\epsfysize=2.2in
\epsffile{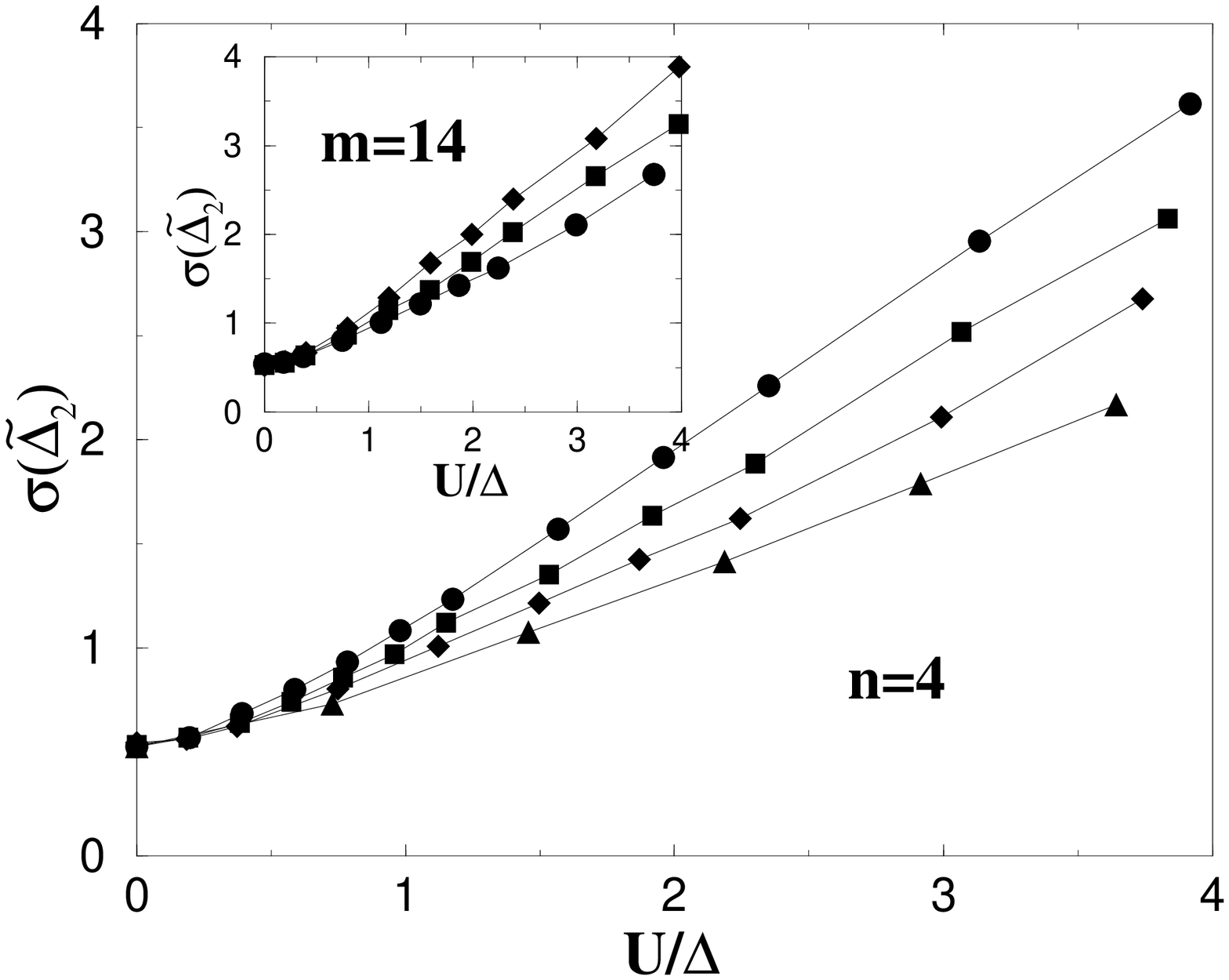}
\epsfxsize=3.3in
\epsfysize=2.2in
\epsffile{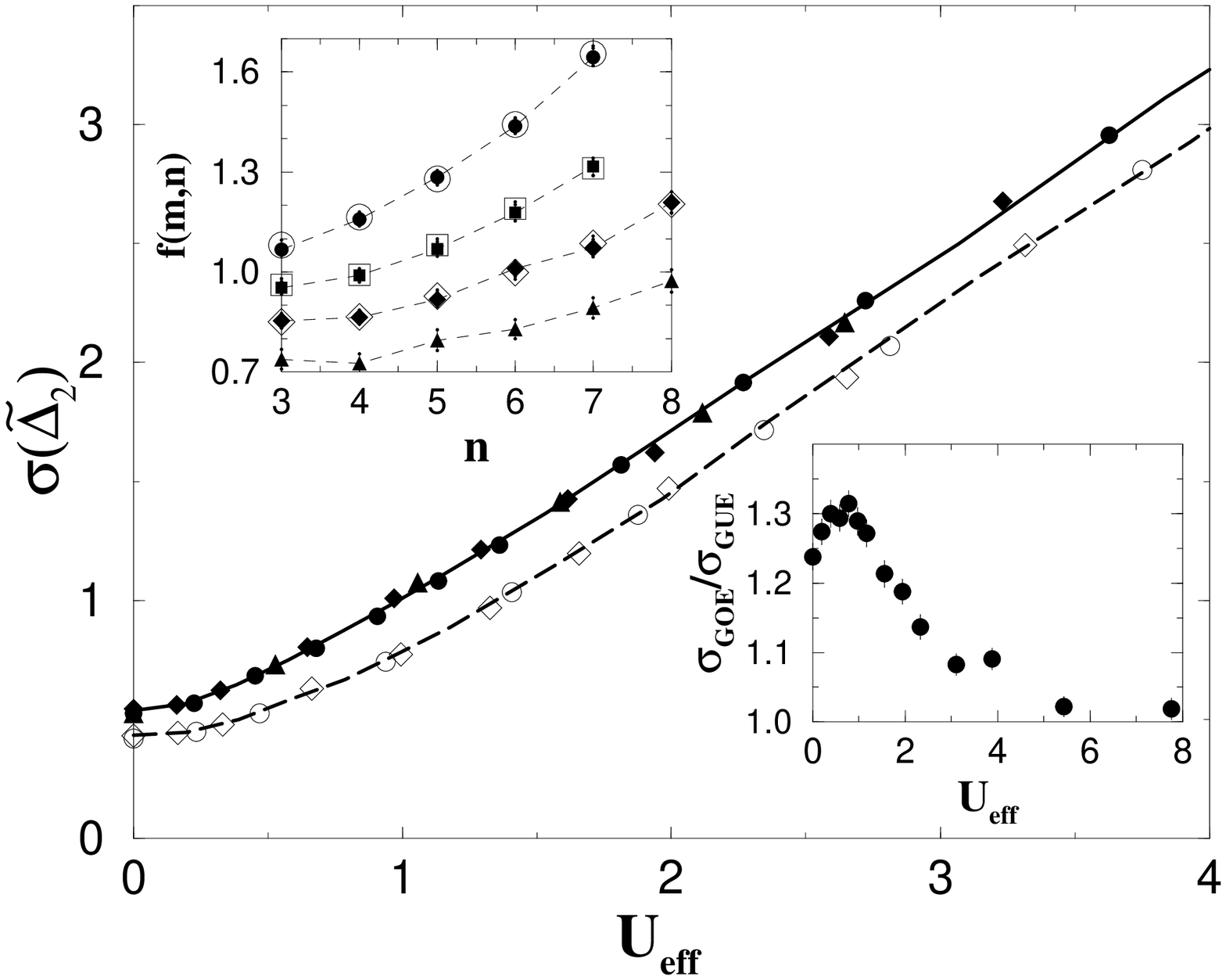}

\caption
{ Top panel: the standard deviation $\sigma(\tilde\Delta_2)$ of the peak
spacing  fluctuations as a function of $U/\Delta$ for the random matrix model
(\protect\ref{Hamiltonian})  with a GOE $h$. Shown are curves for
$n=4$ and $m=10$ (circles), 12 (squares), 14 (diamonds) and 16 (triangles).
Top inset: similar curves for $m=14$ and $n=4$ (circles), 6 (squares) and 8
(diamonds).
Bottom panel: The same curves as in the top panel but
as a function of $U_{\rm
eff}=f(m,n)U/\Delta$. The curves are shown by their corresponding solid symbols
 except for the reference curve ($m=12,n=4$) which is shown by a solid line.
Similar scaled curves ($n=4$ and $m=10,14$)
are shown for the GUE case (open symbols) and compared with the GUE
 reference curve (dashed line).
Left inset: the scaling factors $f(m,n)$ as a function of $n$
for the GOE (solid symbols) and the GUE (open symbols) for
$m=10, 12, 14$ and $16$ (from top to bottom).
Bottom inset: the ratio $\sigma_{\rm GOE}(\tilde\Delta_2) / \sigma_{\rm
GUE}(\tilde\Delta_2)$ versus $U_{\rm eff}$ calculated using
 the $m=12, n=4$ data.
}
\label{fig:sigma}

\end{figure}

 The standard deviation of the
spacing fluctuations $\sigma(\tilde\Delta_2)$ (in units of $\Delta$)
 is shown in the top panel of Fig. \ref{fig:sigma}
vs.  $U/\Delta$ for $n=4$ and several values of $m$. The statistical errors
(due  to finite number of samples) are also estimated but are smaller 
than the size of the symbols.
 At zero interaction we are close to the GOE value of $\approx 0.52$.
$\sigma(\tilde\Delta_2)$  increases slowly and then more rapidly
above $U/\Delta
\sim  0.5$. At strong interactions it is approximately linear in $U/\Delta$. The
top  inset of Fig. \ref{fig:sigma} shows similar curves of the spacing
fluctuations  but for constant number of single-particle states $m=14$ and for
several  values of $n$. The standard deviation curve versus $U/\Delta$
depends on both $m$ and $n$. However upon the linear scaling $U_{\rm eff} =
f(m,n)U/\Delta$  ($f(m,n)$ is a scaling constant) all curves coalesce to a
single universal curve.

To demonstrate the scaling we first choose a
`reference' curve, e.g., $m=12$ and $n=4$, which we determine accurately
using 10,000 realizations at each value of $U/\Delta$. For other values of
$(m,n)$  we use typically $ \sim 1000 - 5000$ realizations and find the scaling
factors  $f(m,n)$ by a least squares fit. The bottom panel of Fig.
\ref{fig:sigma} shows the same curves of the top panel (solid symbols) in
comparison  with the reference curve (solid line), but as a function of the
scaled  parameter $U_{\rm eff}$. The curves scale almost perfectly
within  the statistical errors. Also shown are scaled GUE curves (open
symbols)  for $n=4$ in comparison with the GUE reference curve (dashed line).

 The scaling factor $f(m,n)$ is shown in the left inset of the bottom panel
of  Fig. \ref{fig:sigma} as a function of $n$ for
different  values of $m$ and for both the GOE
(solid symbols with  error bars)
and the GUE (large open symbols) cases. Within the statistical
errors  $f(m,n)$ is independent of the symmetry class, supporting the
universality  of our model.

\begin{figure}
\epsfxsize=3.3in
\epsfysize=2.2 in
\epsffile{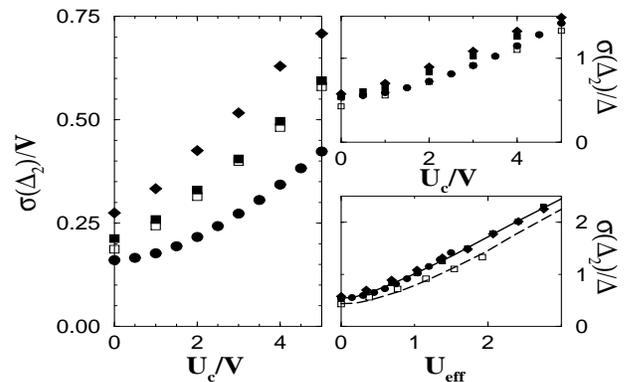}

\caption
{ Left panel: $\sigma(\Delta_2)$ versus $U_c/V$ for a $4\times 5$ cylindrical 
Anderson lattice with Coulomb interactions. Shown are curves
for $\Phi=0$ and a disorder strength of $W=3, 5, 7$ (solid circles, squares and
diamonds, respectively), and for $\Phi =0.15 \Phi_0$ and $W=5$ (open squares).
Right top panel: $\sigma(\Delta_2)/\Delta$
versus $U_c/V$ for the same cases shown on the left panel. Notice
the weak dependence on $W$. Right bottom panel:
$\sigma(\Delta_2)/\Delta$ for
the  Coulomb model (symbols) versus the scaled $U_{\rm eff} = c_0 U_c/V$ in
comparison  with the reference curves of the GOE (solid) and GUE (dashed)
 random matrix model.
}
\label{fig:Coulomb}
\end{figure}

The width of the spacing distribution is larger for the GOE case than for the
GUE  case at any value of $U_{\rm eff}$.
 The bottom right inset of Fig. \ref{fig:sigma} is the ratio $\sigma_{\rm
GOE}(\tilde\Delta_2)/\sigma_{\rm
GUE}(\tilde\Delta_2)$ versus $U_{\rm eff}$, calculated from the reference
curves.   The ratio is $\sim 1.24 \pm 0.02$ for the non-interacting case (in
close  agreement with the RMT value), and depends only weakly 
 on the interaction in the crossover regime $U_{\rm eff} \alt 1$.
This  is consistent with recent measurements \cite{Patel98} which find a ratio
of  $\sim 1.2-1.3$ for semiconductor quantum dots with a gas constant 
of $r_s \sim 1-2$. At stronger interactions the ratio
decreases. At large values of $U$, the two-body Hamiltonian dominates and one
can  ignore the one-body part. Since it is only the latter that distinguishes
between  the conserved and broken time-reversal symmetry cases, the ratio of
the  widths approaches $1$ at strong interactions.

 Next we compare the crossover observed in the RIMM to the results for a
tight-binding Anderson model with cylindrical geometry,
hopping parameter $V=1$, and on-site disorder
with a box distribution of width $W$. Electrons at different
sites interact via a Coulomb interaction whose
strength is $U_c=e^2/a$ over one lattice constant $a$. The standard
deviation of the peak spacing $\sigma(\Delta_2)$ is shown in the
left panel
of  Fig. \ref{fig:Coulomb} versus $U_c/V$ for a $4\times 5$ lattice,
$n=4$ electrons and several values of $W$. The values of $W$
are chosen so that the RMT statistics is approximately satisfied in the
non-interacting case. In the absence of a magnetic field we choose $W=3, 5,
7$.  However, in the presence of a magnetic flux, which we apply inside the
cylinder  and incorporate in the hopping matrix elements in the
 perpendicular direction
($\Phi= 0.15 \Phi_0$), only the $W=5$ case satisfies the spectral
 RMT statistics. We
find that $\sigma_{\rm B=0}(\Delta_2)$ is monotonically increasing
with $W$.
After  rescaling $\sigma$ by the mean level spacing $\Delta$ at the Fermi
energy,  the residual $W$-dependence of $\sigma(\Delta_2)/\Delta$
is rather
weak  (top right panel of Fig. \ref{fig:Coulomb}). The standard deviation
 curves for the Coulomb
model  can be mapped approximately on the random matrix model curve
(bottom  right panel of Fig. \ref{fig:Coulomb}) by defining
 $U_{\rm eff} = c_0 U_c/V$  for some constant $c_0$ that depends
 on the disorder strength and lattice
size.  $c_0$ depends only weakly on $W$.

\begin{figure}

\epsfxsize=3.3in
\epsfysize=2.2in
\epsffile{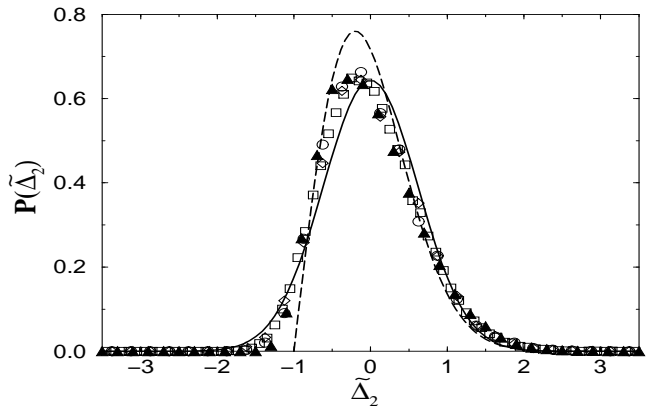}

\caption
{Peak spacing distributions for the random matrix model
(\protect\ref{Hamiltonian}) (where $h$ is a GOE matrix) for
$m=10$, $n=6$, $U/\Delta =0.24$ (open circles); $m=12$, $n=4$, $U/\Delta
=0.33$  (open squares), and $m=14$, $n=4$, $U/\Delta = 0.42$
(open diamonds). In all three cases $U_{\rm eff} = 0.33$.
The solid triangles show the peak spacing distribution of the Coulomb
model for $n=4$ electrons on a $4\times 5$ lattice with $W=5$ and $U_c/V=0.75$.
}
\label{fig:space_dist}

\end{figure}

 The universal aspects of the crossover can also be investigated 
by studying the peak spacing  distributions themselves. In Fig.
 \ref{fig:space_dist} we show the
peak  spacing distributions for three different values of $(m,n)$ but at
constant  $U_{\rm eff} = 0.33 $ (corresponding to three different values of
$U/\Delta$).  We find that all three distributions coincide, indicating that
finite  size effects in the random matrix model are negligible.
 A corresponding distribution for the Coulomb model is
also shown for $U_c/V=0.75$, $W=5$ and $n=4$ and is rather close
 to the random
matrix  model distributions. The deviations seen in the
 Coulomb case may be partly
due to finite size effects that are non-universal; even at $U_c=0$
we  observe deviations from the expected Wigner-Dyson distribution.

 In conclusion, we showed that a random interaction matrix model that
 includes a one-body
part  belonging to one of the standard Gaussian ensembles and a two-body
random  interaction is suitable for studying generic interaction effects on
the  statistics of finite Fermi systems whose single-particle dynamics is
chaotic.  We applied the model to chaotic dots in the Coulomb blockade regime,
where  it describes a crossover of the peak spacing statistics from
a Wigner-Dyson to a Gaussian-like distribution.

This work was supported in part by the U.S. Department of Energy grant
No.\ DE-FG-0291-ER-40608 and the Swiss National Science Foundation. A.W.
acknowledges financial support from the Studienstiftung des deutschen Volkes.

\end{document}